# Students' Responses To Different Representations Of A Vector Addition Question

Jeffrey M. Hawkins*, John R. Thompson*¶, Michael C. Wittmann*¶, Eleanor C. Sayre†, Brian W. Frank*

*Department of Physics and Astronomy, University of Maine, 5709 Bennett Hall, Orono, Maine 04469
¶Maine Center for Research in STEM Education, University of Maine, Orono, ME 04469
†Department of Physics, Wabash College, Box 352, Crawfordsville, IN 47933

**Abstract.** We investigate if the visual representation of vectors can affect which methods students use to add them. We gave students one of four questions with different graphical representations, asking students to add the same two vectors. For students in an algebra-based class the arrangement of the vectors had a statistically significant effect on the vector addition method chosen while the addition or removal of a grid did not.

## INTRODUCTION

Vectors are a ubiquitous part of physics, and even some early mechanics concepts can't be learned without having an understanding of vectors and how they are added and subtracted. It has been shown in several previous studies that students often lack the ability to add vectors [1-3], and that this can lead to student difficulties with other physics topics [4,5].

Vector addition questions can be graphically represented in many ways. Vectors are translationally invariant and can therefore be depicted in infinitely many arrangements. There can also be different visual aides given to the students to help them add or subtract the vectors, including grids, coordinate axis, or angles or magnitudes labeled with specific values. Additionally, questions may consist of any number of vectors, each of which can have any direction or magnitude.

Past research has shown that students use many different methods to add vectors, and that any single student could have multiple methods to choose from [2,6].

The goal of the research presented in this paper is to investigate if there is a link between the way vector addition questions are represented and the methods students choose to add vectors together.

## QUESTION DESIGN

We developed four graphical vector addition questions. Each question has the same two vectors being added, but different graphical representations and features provided. To design these questions we used previous research [1,2,6] to compile a list of prominent graphical vector addition methods used by students. For each method we listed the representational differences we hypothesized would have the strongest effect in getting students to either use or not use that method.

There are several graphical vector addition methods used by students that we considered when designing the questions [6]. The most common of these methods are the *head-to-tail* method and the *components* method. A less common, but still frequently used method is the *bisector* method. In the *head-to-tail* method, students arrange the vectors into a head-to-tail arrangement and then connect the free tail to the free head. In the *components* method students break the vectors into their components and then add the components of both vectors together to get the resultant vector. The *head-to-tail* and *components* methods are both commonly taught as standard methods and can often lead to correct answers. The *bisector* method, which is not taught as a standard method, has many variations. In using the

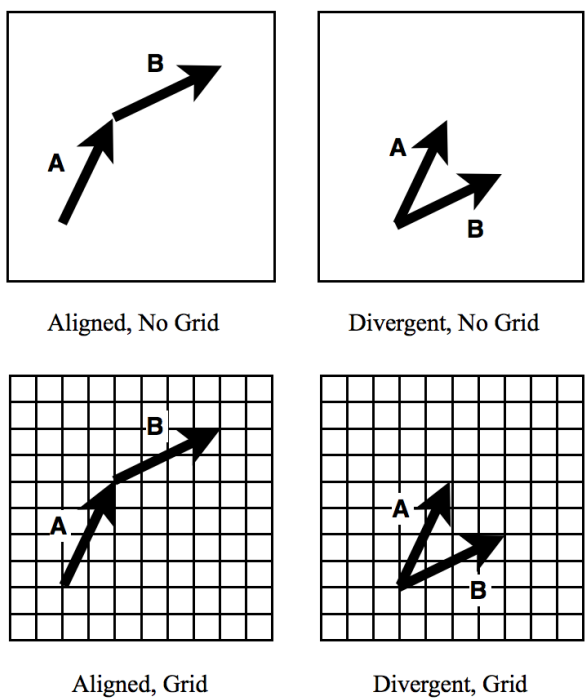

**FIGURE 1.** Students were asked to "Add the two vectors **A** and **B** to get a new vector **R** where **R=A+B**." and explain how they arrived at their answer. The questions have been modified to fit in this paper.

*bisector* method, students determine the direction of the resultant by picking a direction between the two vectors' directions; they then make the length of the resultant somewhere between half the length of the longer vector and the magnitudes of the two vectors added together. Due to the inexactness of the bisector method it can often lead to incorrect answers.

In order to be able to attribute a particular representational change to a resulting shift in method use, we limited ourselves to choosing two different representational changes to make. This gave us the 2x2 grid of question representations shown in Fig. 1.

## Arrangement

We hypothesized a strong effect due to the different arrangement of vectors into either head-to-tail (aligned) or tail-to-tail (divergent) arrangements. By aligning the vectors in a head-to-tail arrangement we are essentially performing the first step of the *head-to-tail* method for students; by aligning them in a tail-to-tail arrangement we are doing the first step of the most common versions of the *bisector* method. This difference in initial representation of the vector pair should cue students to use different addition methods.

## Grid

Our prediction for the second most effective representational change was to add or remove a coordinate grid. Nguyen and Meltzer found that some students who could solve a vector addition question with a grid present could not do so without a grid [2].

Having a grid gives the students numbers to work with and coordinate axes with which to describe the components. We predicted that the use of the *components* method would be less frequent without a grid due to the increased difficulty or ambiguity of using the method. Instead we predicted students would use the *head-to-tail* or *bisector* methods, the two other commonly used methods.

## ADMINISTRATION AND ANALYSIS

The data presented in this paper is from the algebra-based introductory physics course at the University of Maine (UMaine). The questions were given to students in lecture before a five-day break from physics instruction and four-day fall break from school, which occurred between the seventh and eighth weeks of the semester. The students were randomly given either an aligned or divergent arrangement question, without a grid, in the last lecture before the break. Questions with a grid were randomly distributed in recitation on the first day back from break. Each student answered two questions in total, but only one (non-grid) in lecture and one (grid) in recitation.

The reason for giving each student two questions was to increase the number of responses we would receive. In our data analysis we made the assumption that students answered the pre-break (no-grid) and post-break (grid) essentially independently of each other. The questions without a grid were given before the questions with a grid to prevent students from referring to the (provided) grid on the questions without a grid. Possible effects of priming will be discussed later in the paper.

Data were coded for the vector addition methods students used. Because the method students used in their drawings did not always match that described in their written explanations, their drawn explanations were coded separately from their written explanations. The drawn explanations and written explanations were coded into the same set of categories.

The statistical analysis of the data was done using Fisher's exact test. Each of the graphical vector addition methods was tested against the other two methods across the arrangement and grid changes. Fisher's exact test was used instead of the chi-squared test to avoid issues with low cell counts.

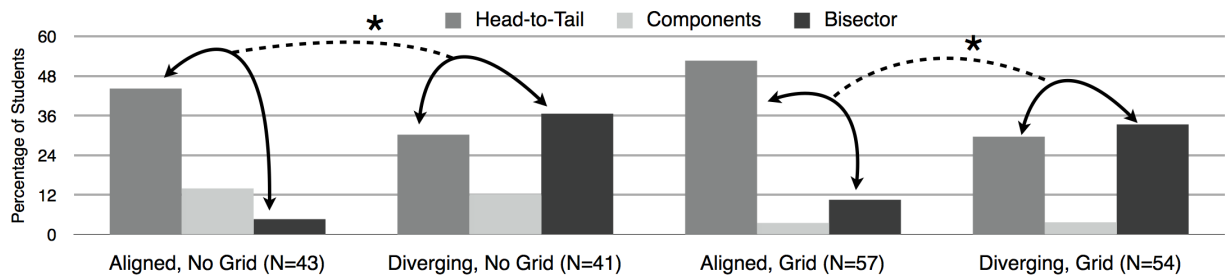

**FIGURE 2.** The percentage of students who used each method in their drawn explanation. Other methods or blanks are not shown. The arrows indicate which methods have significantly* different distributions between particular questions. *Statistically significant, determined by $p < .05$ on Fisher's exact test.

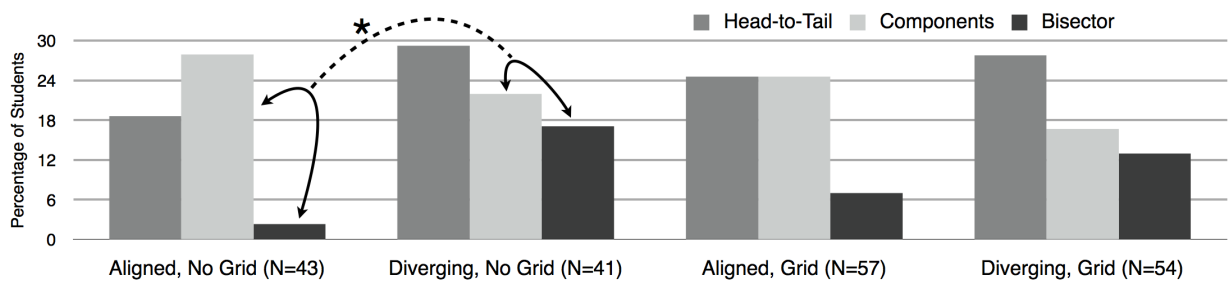

**FIGURE 3.** The percentage of students who used each method in their written explanation. Other methods or blanks are not shown. The arrows indicate which methods have significantly* different distributions between particular questions. *Statistically significant, determined by $p < .05$ on Fisher's exact test.

# FINDINGS: EFFECTS OF REPRESENTATIONAL CHANGES

In the UMaine algebra-based class the changes in the graphical representations led to statistically different distributions of response methods.

## Effects Due To Arrangement

There was a significant difference between the use of the *head-to-tail* and *bisector* methods in students' drawn explanations [Fig. 2] when comparing questions with aligned and divergent arrangements. As predicted, students were more likely to depict the *head-to-tail* method than the *bisector* method on aligned arrangement questions. Students were equally likely to depict the *head-to-tail* method as the *bisector* method on divergent arrangement questions. These differences were statistically significant (Fisher's exact test, $p < .05$) for both the pair of questions with a grid and the pair of questions without a grid.

Our hypothesis—that the arrangement of vectors in a graphical vector addition question affects the relative distribution of the *head-to-tail* method and the *bisector* method—is true only for the students' drawn explanations and not for their written explanations.

There was also a significant difference (Fisher's exact test, $p < .05$) between the use of the *components* and *bisector* methods in students' written explanations [Fig. 3] between questions with aligned and divergent arrangements. This difference in method use was only significant for the pair of questions without a grid. Students were more likely to explain their reasoning by describing the *components* method on questions with an aligned arrangement and about equally as likely to use the *components* method as to use the *bisector* method on questions with a divergent arrangement.

This difference is not one that we had predicted. Further work is necessary to understand why students' explanations would vary in this way, and more generally, why students written explanations do not match their drawn explanations.

## Effects Due To Grid

Our second hypothesis—that use of the *components* method is affected by the presence of coordinate grids—was not confirmed in either the drawn or written explanations [Fig. 2, Fig. 3]. This result is surprising, given the previous results of

Nguyen and Meltzer [1]. We had expected this difference to be of the same order as the difference due to arrangement.

One explanation could be that the students used the same approach they used on the question received before their break. In interviews involving a series of similar vector addition questions [6], we have noted that students often choose one solution method and then use that method on all different representations of the vector addition questions during the interview. This tendency to continue using the same method may be nullifying any effects of the grid in our data. This result could challenge our assumption that the pre-break and post-break questions were answered essentially independently of each other. Ideally we would be able to comment on whether there was such an effect based on the question type received before the break or not. However, there are not enough students in our sample who answered a pair of questions to be able to draw meaningful conclusions.

## CONCLUSION

This study was conducted to see if there are links between the representations of graphical vector addition questions and students responses.

We find that small-scale changes in the representation of graphical vector addition questions can affect the distributions of students' drawn solution methods and written explanations of their solutions.

The arrangement of vectors into head-to-tail (aligned) or tail-to-tail (divergent) arrangements has a significant effect on the way students respond. Students are more likely to use the *head-to-tail* method than the *bisector* method as their drawn explanations when the vectors are arranged in a head-to-tail formation. Students are also more likely to use the *components* method than the *bisector* method as their written explanations on questions with an aligned arrangement but only on questions without a grid.

An identical study in the calculus-based course at UMaine and a similar study, with all four questions randomly distributed simultaneously, at RIT were also carried out, but did not produce insightful data. In each of the calculus-based classes at both RIT and UMaine at least 85% of the responses received were correct, compared to 50% in the algebra-based class. The students who answered the questions correctly mostly did so using the same method and no significant differences in response patterns were seen.

This study did not find an expected link between grid presence and response method. The lack of an effect from adding a grid may be due to priming.

In a previous study [6], priming appeared to be a prominent factor in students' choice of vector addition method. Priming may have also played a significant role in students' method choice in this study, but this time with the priming occurring over a much larger time scale.

## ACKNOWLEDGMENTS

We would like to thank the members of the UMaine Physics Education Research Laboratory, especially Dr. David Clark; Dr. Scott Franklin (RIT), Frank Dudish (UMaine), Jessica Clark (RIT), and the instructors of the introductory physics classes at RIT. This work was supported in part by NSF Grant #DRL-0633951.